\documentstyle[prl,aps,multicol,epsf]{revtex}

\begin{document}

\title{Effects of Spin-Orbit Interactions on Tunneling via Discrete 
Energy Levels in Metal Nanoparticles}

\author{D.~G.~Salinas, S.~Gu\'eron, and D.~C.~Ralph}
\address{Laboratory of Atomic and Solid State Physics, Cornell 
University, Ithaca, NY 14853}
\author{C.~T.~Black}
\address{IBM Thomas. J. Watson Research Center, Yorktown 
Heights, NY 10598}
\author{M.~Tinkham}
\address{Department of Physics and Division of Applied Sciences, 
Harvard University, Cambridge, MA 02138}
\date{\today}
\maketitle

\begin{abstract}
The presence of spin-orbit scattering within an aluminum 
nanoparticle affects measurements of the discrete energy levels 
within the particle by (1) reducing the effective $g$-factor below 
the free-electron value of 2, (2) causing avoided crossings as a 
function of magnetic field between predominantly-spin-up and 
predominantly-spin-down levels, and (3) introducing 
magnetic-field-dependent changes in the amount of current transported by 
the tunneling resonances.  All three effects can be understood in a 
unified fashion by considering a simple Hamiltonian.  Spin-orbit 
scattering from 4\% gold impurities in superconducting aluminum 
nanoparticles produces no dramatic effect on the superconducting 
gap at zero magnetic field, but we argue that it does modify the 
nature of the superconducting transition in a magnetic field.
\end{abstract}

\pacs{PACS numbers:  71.70.Ej, 73.20.Dx, 74.80.Bj}

\begin{multicols} {2}
\narrowtext

\centerline{I.  Introduction}

\vskip 12pt
	For decades, systematic studies of the quantum-mechanical 
energy levels of atoms and atomic nuclei have provided an 
understanding of the forces governing these systems.  Recently, it 
has also become possible to measure the discrete ``electrons-in-a-box''
energy levels within semiconductor quantum dots and metal 
nanoparticles \cite{Ashoori,Kastner,rbt1}.  Experiments have 
shown that different classes of forces and interactions acting on the 
electrons inside these materials affect the level spectra in 
distinguishable ways.  Therefore, just as in atomic and nuclear 
physics, the discrete spectra in these condensed matter systems can 
provide a tool for understanding the interactions which influence 
electronic structure, uncovering effects that are not clearly visible 
if the individual quantum levels in the system cannot be resolved.  
The consequences of superconducting pairing interactions 
\cite{cbt,rbt2} and more general electron-electron interactions 
\cite{Agam1,Agam2} have previously been analyzed for the case 
of aluminum nanoparticles.  In this report, we discuss spin-orbit 
(S-O) interactions, resulting both from accidental defects in the Al 
nanoparticles and from gold dopants.  We examine how S-O 
scattering affects both the energies of the quantum levels and the 
amount of tunnel current which may be carried by each state.  We 
find that the magnetic-field dependence of these quantities may be 
understood in a unified fashion within a simple model.  The effects 
of S-O scattering on the superconducting properties of an 
aluminum nanoparticle are also discussed.

	The study of S-O scattering within metals has a long 
history.  The metal samples of the types traditionally examined are 
large enough that the electronic states effectively form a 
continuum.  In this case, the quantity of primary experimental 
interest in S-O studies is the rate at which S-O interactions cause 
the spin of an electron assumed to be initially in a pure spin-up or 
spin-down state to be scattered into continuum states with opposite 
spin.  This rate can be measured using weak localization 
experiments for disordered metal samples \cite{Bergmann} or, 
alternatively, by analyzing the form of the spin-dependent density 
of states determined by tunneling between thin superconducting 
films in a parallel magnetic field \cite{Meservey}.  S-O 
interactions are of fundamental theoretical importance because 
their presence changes the symmetry properties of the 
Hamiltonian.  For instance, the statistics of the energy levels in 
chaotic time-reversal-symmetric quantum dots are predicted to 
change from the orthogonal distribution in the absence of S-O 
scattering to the symplectic distribution for a strong S-O 
interaction, with a corresponding increase in the strength of the 
effective energy-level repulsion \cite{Halperin,Perenboom}.  
Perhaps the most dramatic consequences of S-O coupling in metals 
occur in ferromagnets, since the S-O interaction underlies the 
phenomena of magnetic anisotropy and the anomalous Hall effect.

	An analysis of the effects of S-O interactions in metal 
nanoparticles requires a somewhat different viewpoint than for 
larger devices with a continuum density of states.  Considering 
basic symmetries, the Hamiltonian operator describing electrons 
within a metal sample does not commute with the components of 
the total electronic spin operator in the presence of the S-O 
interaction.  This means that it is not possible to construct a set of 
basis states which are simultaneously eigenstates of both the 
energy and $S_z$.  The discrete energy eigenstates, through which 
electron tunneling occurs in a metal nanoparticle, will thus 
necessarily be linear superpositions of pure spin-up and pure 
spin-down states, with the extent of admixture determined by the 
magnitude of S-O matrix elements.  Because these discrete energy 
eigenstates defined in the presence of the S-O interaction are in 
fact well-defined energy eigenstates, the S-O interaction does not 
lead to any decrease in their lifetime.  For this reason, the 
experimental quantities of interest in this paper will not be 
scattering rates, but rather shifts in the energies of the electronic 
states and changes in the tunneling current carried by the states 
\cite{estimate}.  Both of these quantities are affected by the extent 
of admixture of spin-up and spin-down components within the 
energy eigenstates.  An initial analysis of some of the results we 
will discuss has appeared previously \cite{Black}.

	The measurements we describe were performed using 
tunneling devices containing an Al particle less than 10~nm in 
diameter, connected to Al electrodes via aluminum oxide tunnel 
junctions.  A device schematic is shown inset to Fig.~1(a).  The 
fabrication steps have been described in detail previously 
\cite{rbt1}.  An aluminum electrode is first deposited on one side 
of an insulating silicon nitride membrane containing a 10-nm-scale 
through-hole.  The Al is oxidized to form a nm-scale tunnel 
junction near the base of the hole.  A layer of Al nanoparticles is 
then formed on the other side of the membrane by depositing 
2.5~nm of Al, which balls up into small particles due to surface 
tension.  In some of the devices described in this paper, the Al 
evaporation for the particles was interrupted half-way through and 
a thin layer of gold was deposited to give roughly a 4\% (atomic) 
dose of Au inside the nanoparticle.  Since Al and Au are 
sufficiently miscible to form several intermetallic compounds 
\cite{Hansen}, and both have significant surface mobilities on the 
nm length scale, we expect that the two types of atoms will be 
intermixed.  When the nanoparticle deposition is complete, their 
surfaces are oxidized to form tunnel junctions, and a thick 
aluminum film is deposited as a second electrode.  Devices in 
which tunneling occurs via a single nanoparticle joining the two 
leads are selected based on the measurement of a 
``Coulomb-staircase" current-voltage curve (Fig.~1(a)).

\vskip 24pt

\centerline{II.  Effects of Spin-Orbit Interactions on Discrete 
States}

\vskip 12pt
	Tunneling spectra of the discrete energy levels are shown in 
Fig.~1(b), at different values of the applied magnetic field, for an 
Al particle in which we will identify the presence of spin-orbit 
scattering.  This particle is nominally pure Al, but we have also 
observed all the features that we will ascribe to S-O scattering in 
Au-doped particles.  We speculate that the source of the S-O 
scattering in the nominally pure Al particle is an unintended de-
\linebreak
\begin{figure}
\vspace{-1.1cm}
\begin{center}
\leavevmode
\epsfxsize=2.5in
\epsfbox{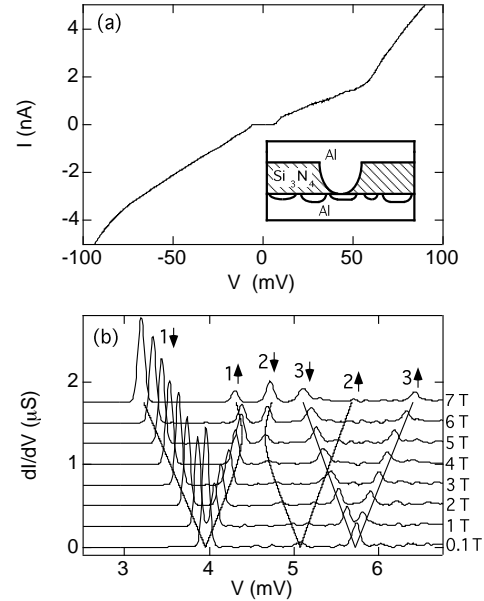}
\end{center}
\caption{
(a) Large scale Coulomb-staircase curve for a tunneling 
device containing a nm-scale Al particle at $T\!=\!50$ mK.  
Inset:  Cross-sectional device schematic.  (b)  Tunneling 
spectrum of discrete state resonances in the same sample, 
for a range of applied magnetic fields, at $T\!=\!50$ mK.  
The curves are offset in $dI/dV$ for visibility.  Orbital state 
\#2 gives small but visible resonances at low $B$.  Small 
changes in offset charge occurred between the 0.1 and 
1~Tesla scans and between the 6 and 7~Tesla scans, 
shifting peak positions.  The 0.1 and 7~Tesla scans have 
therefore been shifted along the voltage axis, to give the 
best fit to a linear dependence for peak $1\downarrow$.  
The lines tracing under peaks are guides to the eye.}
\end{figure}
\noindent
fect or impurity.  Each peak in the $dI/dV$ vs.~voltage spectrum 
corresponds to the threshold for electron tunneling via a different 
quantum-mechanical state in the particle, each with the same 
number of electrons (either one more or one less than in the 
$V\!=\!0$ ground state of the particle).  In order to convert from 
the voltage scale to the true energy within the nanoparticle, it is 
necessary to determine the ratio of the capacitances of the particle 
to the two electrodes, $C_1/C_2$.  This is measured most 
accurately either by comparing the positions of the tunneling peaks 
due to the same state at opposite signs of bias voltage, or by 
measuring shifts in peak positions for superconducting vs.~normal-state 
electrodes \cite{rbt1}.  The conversion factor from voltage to 
energy for the data of Fig.~1(b) is $eC_1/(C_1+C_2) = e 
(0.53\pm0.01)$.  A rough estimate of the volume of the particle 
can be made based on the capacitances of the particle, determined 
from the spacing between steps in the Coulomb-staircase curve, 
$\Delta$$V\!=\!e/C_{smaller}\!=\!78$ mV.  Together with the known 
capacitance per unit area of oxidized aluminum tunnel junctions, 
$\sim50~fF/\mu$$m^2$ \cite{Lu}, and assuming a particle shape 
that is approximately a hemisphere, we estimate a particle radius of 
approximately $r\!=\!3$ nm for this device.

	The peaks in Fig.~1(b) have many features qualitatively 
similar to previous studies of tunneling resonances in pure Al.  As 
the applied magnetic field ($B$), applied parallel to the plane of 
the $Si_3N_4$ membrane in the device, is increased from low-field 
values, each peak splits in two, and the energy difference between 
these pairs increases linearly with $B$ at low $B$ 
(Fig.~2(a)) \cite{Zerob}.  This can be understood as Zeeman splitting of the 
energies of the predominantly spin-up and spin-down states 
associated with each orbital eigenstate.  The observation of 
tunneling via both of the Zeeman-split states for the lowest-energy 
tunneling state ($\#1$) indicates that the tunneling transition 
corresponds to a change from an even number to an odd number of 
electrons within the nanoparticle \cite{rbt1}.  Within the 
uncertainties of the measurement, the splitting is symmetric around 
the low-field resonance energy, with little shift up or down for the 
average of the Zeeman-split peaks \cite{Upbound}.  This indicates 
that the effect of $B$ on the orbital component of the electronic 
energy is much weaker than on the spin component.  This is not 
surprising, due to the particle's small size and disorder.  Because 
any real nanoparticle will not have a spherical shape or a smooth 
surface, the orbital angular momentum of the eigenstates will be 
quenched to zero in the absence of an applied field.  In this 
situation, the correlation scale which describes the effect of the 
magnetic field on the energy eigenstates is expected to be on the 
order of $\Phi_0\sqrt{\delta/E_{Th}}/r^2$, where $\Phi_0$ is the 
flux quantum, $\delta$ is the mean level spacing, and 
$E_{Th}\approx\hbar v_F/(2r)$ is the Thouless energy scale for a 
ballistic sample \cite{Falko}.  For a particle with radius 3~nm the 
expected field correlation scale is approximately 30~Tesla.  Since 
this is much larger than the fields of interest in our experiment, 
throughout the paper we will assume that the effect of $B$ on the 
orbital eigenstates within the particle is negligible, so as to 
concentrate on spin effects.

	There are at least 3 features of the data in Figs.~1(b) and 
2(a) that differ from typical Al particles.  Firstly, let us define an 
effective $g$ factor such that the energy splitting between Zeeman-split
states is $\Delta\!E=g_{\rm{eff}}\mu_{B}B$ (to linear order in 
$B$), where $\mu_{B}$ is the Bohr magneton.  In over 80\% of 
the nominally pure Al samples we have examined previously, 
$g_{\rm{eff}}\!=\!2\pm0.05$, which is as expected, because S-O 
scattering is negligible in pure Al, and the free electron $g$-factor 
should apply \cite{Halperin}.  In the sample in question, however, 
$g_{\rm{eff}}$ is significantly less, and it varies from peak to peak: 
$g_{\rm{eff}}\!=\!1.84\pm0.03$, $1.68\pm0.08$, and $1.76\pm0.05$ 
for the 3 resonances in Fig.~2(a).  The second difference between 
this sample and past measurements concerns level crossings.  In 
pure Al particles with $g$-factors approximately equal to 2 we 
have not observed departures from linear Zeeman splittings when 
spin-up and spin-down levels corresponding to different orbital 
states cross as a function of $B$.  For a sample without S-O 
scattering, this must be the case, for then there is no coupling 
between spin-up and spin-down states in
\linebreak
\begin{figure}
\vspace{-1.1cm}
\begin{center}
\leavevmode
\epsfxsize=2.5in
\epsfbox{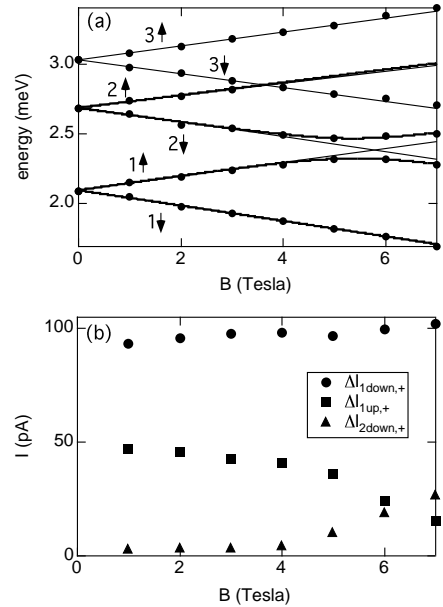}
\end{center}
\caption{
(a) Energies of the discrete electronic states within the 
nanoparticle of Fig.~1, calculated by multiplying the 
voltage positions of the resonances by the capacitance ratio 
$eC_1/(C_1+C_2) = e$ 0.53.  The thin lines are extensions of 
the low-field linear dependence of the energies on $B$.  
Heavy lines show the result of the spin-orbit interaction 
model, describing the avoided crossing between levels 
$1\uparrow$ and $2\downarrow$.  (b) Magnitude of the 
current increments contributed by each of the first three 
resonances for positive voltage bias.  (Equal to the area 
under the peaks in Fig.~1(b).)  Note the crossover in 
magnitude for the current increments associated with states 
$1\uparrow$ and $2\downarrow$.}
\end{figure}
\noindent
the Hamiltonian.  In 
contrast, the Zeeman splittings of the first two orbital states shown 
in Figs.~1(b) and 2(a) show a clear departure from linear behavior, 
because the upward-trending level from the first state 
($1\uparrow$) undergoes an avoided crossing with the downward-trending 
level from the second state ($2\downarrow$).  The third 
difference is that the amplitudes of the resonances for the sample in 
question show unusual features.  Whereas in most Al particles the 
amplitudes of the resonances do not display any significant 
$B$-dependence, here the amplitudes of the two levels undergoing the 
avoided crossing change dramatically, with the higher-amplitude 
resonance shrinking and the smaller-amplitude resonance growing 
in the avoided crossing region.  The amount of current contributed 
by each resonance is plotted in Fig.~2(b), where it can be seen that 
the sum of the current increments contributed by these two 
resonances is approximately constant.

	All of these features can be understood by considering the 
Hamiltonian of the electrons in the presence of S-O scattering.  
The theory behind the physics of the $g$-factor has been considered 
previously \cite{Sone,Halperin}.  Let us write the Hamiltonian in 
zero magnetic field as
\begin{equation}
H=H_{0}+H_{SO},
\end{equation}
where $H_{SO}$ contains the terms that couple spin-up states to 
spin-down, and $H_0$ describes all the spin-independent forces 
and interactions.  We will neglect the effect of the magnetic field 
on electron orbits, and assume that the sample contains no 
magnetic impurities.  Let $|A_{n\uparrow}\rangle$ and 
$|A_{n\downarrow}\rangle$  represent the unperturbed 
eigenstates of $H_0$.  Then, performing perturbation theory to 
lowest order in $H_{SO}$, the (not normalized) eigenstates of the 
full Hamiltonian have the form
\begin{equation}
|\Phi_{n\rm{``}\uparrow\it{''}}\rangle = |A_{n\uparrow} \rangle + 
\sum_{m \ne n}{{\langle A_{m\downarrow}
|\!H_{SO}\!|A_{n\uparrow}\rangle |
A_{m\downarrow}\rangle}\over {E_n-E_m}}.
\end{equation}
The spin-orbit interaction causes the eigenstates to consist of a 
linear superposition of spin-up and spin-down states; hence the 
notation ``$\uparrow$'' reflects that the eigenstate can be 
considered at most predominantly spin-up.  The effective 
$g$-factor for state $n$ may be written \cite{Halperin}
\begin{equation}
 \matrix{ \displaystyle g_{\rm{eff},n}
\equiv 2{{|\langle\Phi_{n\rm{``}\uparrow\it{''}}| \sigma_z |
\Phi_{n\rm{``}\uparrow\it{''}}\rangle |} \over 
{\langle\Phi_{n\rm{``}\uparrow\it{''}} | 
\Phi_{n\rm{``}\uparrow\it{''}}\rangle}} \cr
\displaystyle \hfill = 2 \Bigl(1-2\sum_{m \ne n} 
{{|\langle A_{m\downarrow}|\!H_{SO}\!|A_{n\uparrow}\rangle |^2} 
\over {(E_n-E_m)^2}} \Bigr) .}
\end{equation}
(Evaluating this expression for the ``$\downarrow$'' state gives the 
same answer.)  The meaning of Eq.~(3) is that $g_{\rm{eff}}$ is 
reduced below the free-electron value of 2 by an amount 
determined by the extent to which S-O matrix elements couple the 
state $n$ to other states $m$ of opposite spin.  Because the energy 
eigenstates are no longer purely spin-up or spin-down in the 
presence of spin-orbit interactions, they respond more weakly to an 
applied magnetic field than pure-spin states.  Next consider the 
nature of the matrix elements 
$|\langle A_{m\downarrow} |\!H_{SO}\!|A_{n\uparrow}\rangle |^2$.  
Due to the chaotic and strongly fluctuating character of the 
wavefunctions in a metallic nanoparticle \cite{Agam1}, the 
magnitudes of these factors will be strongly varying for different 
values of $m$ and $n$, depending on the details of the wavefunction 
overlaps at the positions of the S-O scattering defects.  Therefore, 
from Eq.~(3), it can be seen that different energy levels in the same 
sample may have different values of $g_{\rm{eff}}$, as we observe.  
Because of the form of the denominator in the second term of 
Eq.~(3), we can also expect that matrix elements which couple 
eigenstates nearby in energy will produce the strongest influence 
on $g_{\rm{eff}}$.  We will demonstrate an example of this below.

	Let us now begin to analyze the variations in the level 
energies and the currents carried by the particular levels displayed 
in Fig.~2.  To do this we will write explicitly the form of the 
effective Hamiltonian matrix for just the 4 energy levels associated 
with the first two orbital states, which we label as 
$|a\downarrow\rangle$, $|a\uparrow\rangle$, 
$|b\downarrow\rangle$, and $|b\uparrow\rangle$.  The most 
convenient set of basis states are those which diagonalize the 
spin-independent part of the Hamiltonian, $H_0$, together with all of 
$H_{SO}$ except that term which couples states $|a\rangle$ and 
$|b\rangle$ to each other.  (With this choice, the basis states are 
already not purely spin-up or spin-down, so the arrows should 
henceforth be understood to mean predominantly spin-up or 
predominantly spin-down.)  The S-O interaction is invariant upon 
time reversal.  The most general Hamiltonian satisfying this 
symmetry, including both ordinary potential scattering and S-O 
scattering, and describing two Kramers' doublets in the absence of 
an applied magnetic field is (with the above basis choice) 
represented by the matrix \cite{Perenboom}:
\begin{equation}
H=\left(\matrix{E_{a\downarrow}&0&d&c\cr
0&E_{a\uparrow}&-c^*&d^*\cr
d^*&-c&E_{b\downarrow}&0\cr
c^*&d&0&E_{b\uparrow}}\right).
\end{equation}
The placement of the zero elements and the arrangement of the 
elements involving $c$ and $d$ are required so that the Kramers' 
doublets are in fact degenerate at $B\!=\!0$.  The matrix element 
$d\!=\!\langle$$a\downarrow$$|H_{SO}|b\downarrow\rangle$ couples 
states of the same spin, so that it is equivalent to ordinary potential 
scattering for our purposes.  Without loss of generality, we can 
pick the orbital basis states $|a\rangle$ and $|b\rangle$ so that 
$d\!=\!0$.  We identify 
$c\!=\!\langle$$a\downarrow$$|H_{SO}|b\uparrow\rangle$.  Because 
we are assuming that the orbital states are not modified by a 
magnetic field, we take the matrix element $c$ to be independent 
of $B$.  The only $B$-dependence then left in the problem is due 
to the influence of the Zeeman energies in the diagonal terms of 
the Hamiltonian.  We write these Zeeman energies by including 
effective $g$-factors, $g_{\rm{eff}}^{\prime}$, for the spin and, simply for 
convenience in the fitting, we also allow a linear term 
$g_{orb}\mu_{B}B$ (where $\mu_{B}$ is the Bohr magneton) to 
model any shift in the average energy of the Zeeman-split pairs.  
(We will see that the fits give $g_{orb}\!\approx\!0$.)  With these 
assumptions, the diagonal terms as a function of $B$ are 
\begin{equation}
\matrix{\displaystyle E_{a\uparrow,\downarrow}=E_a + 
(g_{orb,a}\pm{{g_{\rm{eff},a}^{\prime}} \over {2}} ) \mu_{B} B\cr
\displaystyle E_{b\uparrow,\downarrow}=E_b + 
(g_{orb,b}\pm{{g_{\rm{eff},b}^{\prime}} 
\over {2}} ) \mu_{B} B.}
\end{equation}
The terms $g_{\rm{eff},a}^{\prime}$  and $g_{\rm{eff},b}^{\prime}$ 
must take into 
account 
the S-O coupling of state $|a\rangle$ or $|b\rangle$ to all states 
except each other, so that these terms will not be equal to 2.  
Instead, from Eq.~(3), we should expect that 
$g_{\rm{eff},a}^{\prime}$ and 
$g_{\rm{eff},b}^{\prime}$ will be related to the 
total effective $g$ factor by the 
relationship
\begin{equation}
g_{\rm{eff}}^{\prime} = g_{\rm{eff}} + 4 
{{|\langle a\downarrow |H_{SO}|b\uparrow \rangle |^2} \over {(E_a - 
E_b)^2}}.
\end{equation}
With $d\!=\!0$, Eq.~(4) gives a very simple Hamiltonian, 
consisting of two separate 2-by-2 matrices coupling 
$|a\downarrow\rangle$ to $|b\uparrow\rangle$, and 
$|a\uparrow\rangle$ to $|b\downarrow\rangle$.

	For a weak S-O interaction, 
$|\langle a \downarrow$$|H_{SO}| b \uparrow \rangle |\ll E_b-E_a$, 
the effects of the 
interaction are easy to understand.  Away from any degeneracies 
among the diagonal terms, the energy eigenvalues will be 
approximately equal to the diagonal terms, except for a shift in the 
effective $g$-factor.  When the Zeeman energies are such that two 
diagonal energies approach degeneracy, they will exhibit a simple 
avoided crossing of magnitude equal to 
$2|\langle$$a \downarrow$$|H_{SO}|b \uparrow \rangle$$|$, 
because this term couples the two 
states.  Solving the Hamiltonian explicitly (with $d\!=\!0$), the 
model produces an excellent fit for the $B$-dependence of the 
measured levels (Fig.~2(a)), with the parameters $|\langle a 
\downarrow$$|H_{SO}|b \uparrow \rangle| = 73\pm4~\mu$$eV$, 
$g_{\rm{eff},a}^{\prime} = 1.90\pm0.04$, 
$g_{\rm{eff},b}^{\prime} = 1.74\pm0.04$, 
$g_{orb,a} = -0.03\pm0.04$, and $g_{orb,b} = -0.10\pm0.06$.  
The difference between the directly measured $g$-values 
$g_{\rm{eff},1} = 1.84\pm0.03$, $g_{\rm{eff},2} = 
1.68\pm0.08$ on the one 
hand and the fitting terms $g_{\rm{eff},a}^{\prime}$, 
$g_{\rm{eff},b}^{\prime}$ on 
the other 
is consistent with Eq.~(6), since $4|\langle a \downarrow 
|H_{SO}|b \uparrow \rangle|^2 / (E_a - E_b)^2 = 0.06$.  From this 
we can see that the S-O coupling between states $|a\rangle$ and 
$|b\rangle$ contributes approximately 40\% of the reduction from 
$g_{\rm{eff}}\!=\!2$ for the orbital state $1$, and 20\% for state $2$.  
S-O coupling to other states must account for the remainder.  The 
fact that we do not have the sensitivity to resolve any avoided 
crossing between the states $2\uparrow$ and $3\downarrow$ 
(Fig.~2(a)) indicates that the S-O matrix element coupling these 
states is smaller than $\langle a \downarrow$$|H_{SO}|b \uparrow 
\rangle$.

	The changes in the amount of current carried by the 
resonances (Fig.~2(b)) can be understood by examining the 
manner in which the energy eigenstates are composed of linear 
superpositions of basis states.  Consider the two energy eigenstates 
($|lower \rangle$ and $|upper \rangle$) formed from superpositions 
of the avoided-crossing basis states $|a\uparrow\rangle$ and 
$|b\downarrow\rangle$.  By diagonalizing the Hamiltonian 
(Eq.~(4) with $d\!=\!0$), it is simple to demonstrate that these have 
the form
\begin{equation}
\matrix{|lower \rangle = \gamma(B) |a\uparrow \rangle + \eta(B) |b 
\downarrow \rangle\cr
|upper \rangle = -\eta^*(B) |a\uparrow \rangle + \gamma^*(B) |b 
\downarrow \rangle}
\end{equation}
where the coefficients $\gamma(B)$ and $\eta(B)$ depend on $B$ 
as shown in Fig.~3(a).  The key point is that, as the magnetic field 
is varied in the avoided crossing region, the relative contributions 
of $|a\uparrow\rangle$ and $|b\downarrow\rangle$ to each 
eigenstate will change, and consequently the tunneling currents can 
be altered.  This simple conclusion will be the topic of the next 
two pages of discussion.  The reason for an extended analysis is 
that the magnitudes of the currents are determined by a process of 
sequential tunneling across the two tunnel junctions in the device, 
so that the measured current values are not simply a function of a 
the tunneling rate into an individual energy eigenstate.  Instead, the 
current will be affected by all energetically-allowed transitions 
within the device.  In order to deal with this complication, the plan 
of our discussion is that we will focus first on the bare tunneling 
rates  $\Gamma$$_{L,lower}$, $\Gamma$$_{L,upper}$, 
$\Gamma$$_{R,lower}$, and $\Gamma$$_{R,upper}$ for tunneling of 
an electron between the energy eigenstates ($|lower \rangle$ and 
$|upper \rangle$) and the left (L) and right (R) electrodes.  Later 
we will examine two different limits for calculating the total 
current through the device in terms of these bare tunneling rates.  
In either case we will see that, despite the complications,
\linebreak
\begin{figure}
\vspace{-1.1cm}
\begin{center}
\leavevmode
\epsfxsize=2.8in
\epsfbox{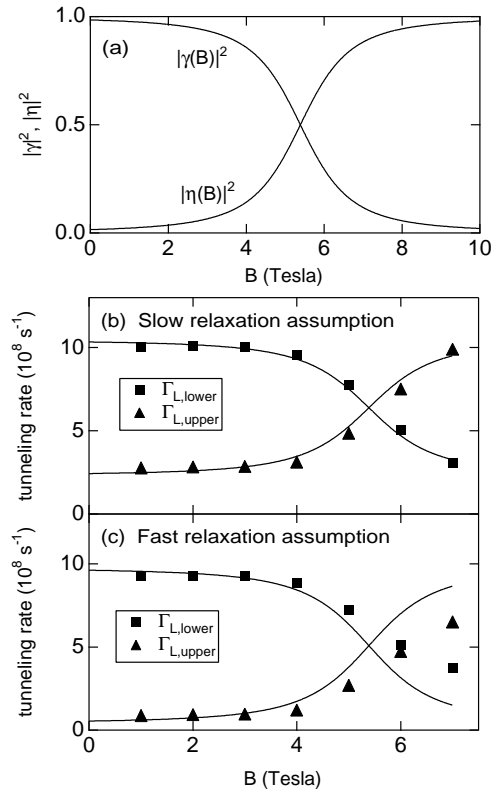}
\end{center}
\caption{
(a)  Magnetic-field dependence of the coefficients in 
Eq.~(9), for the superposition of predominantly spin-up and 
spin-down basis states occurring in the avoided crossing of 
levels $1\uparrow$ and $2\downarrow$.  (b) and (c)  
Markers: Tunneling rates for the energy eigenstates in the 
avoided crossing region, estimated as discussed in the text 
using either Eq.~(11) or (12), assuming that the relaxation 
rate of non-equilibrium excitations within the particle is 
either slower or faster than the tunneling rates.  Lines: Fits 
using the predictions of the spin-orbit Hamiltonian.  
Regardless of the energy-relaxation rate, the magnetic-field 
dependent changes in the currents flowing via the levels in 
the avoided crossing region (Fig.~2(b)) can be explained 
qualitatively by the change in tunneling rates expected from 
the S-O Hamiltonian.}
\end{figure}
\noindent
the 
changes in the total current carried by a particular
tunneling 
resonance as a function of magnetic field can be related to the 
changing composition of the energy eigenstates  in the 
avoided-crossing region (Eq.~(7)).

	For the high-resistance barriers used in the experiment, the 
bare tunneling rates between either of the electrodes and energy 
levels in the nanoparticle can be written in terms of matrix 
elements of a tunneling Hamiltonian $H_T$ which couples states in 
the electrodes to the energy eigenstates.  Since tunneling of a 
spin-up electron from the electrode is necessarily incoherent with 
respect to tunneling of a spin-down electron, we have for the left 
junction (for the right junction the equations are similar):
\begin{equation}
\matrix{\displaystyle  \Gamma_{L,lower}
= {{2\pi} \over {\hbar}} \sum_{\psi in 
left \atop electrode} \Bigl\{ | \langle \psi_{electrode,\uparrow} | 
H_T | lower \rangle |^2 \cr
\displaystyle \hfill + | \langle \psi_{electrode,\downarrow} | 
H_T | lower \rangle |^2 \Bigr\} \cr
\displaystyle \Gamma_{L,upper}
= {{2\pi} \over {\hbar}} \sum_{\psi in left \atop 
electrode} \Bigl\{ | \langle \psi_{electrode,\uparrow} | H_T | upper 
\rangle |^2 \cr
\displaystyle \hfill + | \langle \psi_{electrode,\downarrow} | H_T | upper 
\rangle |^2 \Bigr\}.}
\end{equation}
These expressions can be given in a more illuminating form by 
writing $|lower \rangle$ and $|upper \rangle$ explicitly as linear 
superpositions of the basis states $|a\uparrow\rangle$ and 
$|b\downarrow\rangle$ (as in Eq.~(7)).  The tunneling rates 
become
\begin{equation}
\matrix{ \Gamma$$_{L,lower} = |\gamma(B)|^2 
\Gamma$$_{L,a\uparrow} + |\eta(B)|^2 \Gamma$$_{L,b\downarrow} 
\cr
\Gamma$$_{L,upper} = |\eta(B)|^2 \Gamma$$_{L,a\uparrow} + 
|\gamma(B)|^2 \Gamma$$_{L,b\downarrow}}
\end{equation}
where the $B$-independent tunnel-coupling strengths for the basis 
states are
\begin{equation}
\matrix {\Gamma$$_{L,a\uparrow} = {{2\pi} \over {\hbar}} 
\sum_{\psi} |\langle \psi_{electrode,\uparrow}| H_T | a\uparrow 
\rangle |^2 \cr
\Gamma$$_{L,b\downarrow} = {{2\pi} \over {\hbar}} \sum_{\psi} 
|\langle \psi_{electrode,\downarrow}| H_T | b\downarrow \rangle 
|^2.}
\end{equation}
For values of $B$ well below the avoided crossing range, we have
$\gamma(B) \approx 1$, $\eta(B) \approx 0$, and the rates for 
tunneling into the energy eigenstates are equal to 
$\Gamma$$_{L,a\uparrow}$ and $\Gamma$$_{L,b\downarrow}$.  
Since in this regime the total current passing through the 
$|b\rangle$ resonances (orbital state \#2 in Figs.~1(b), 2(b)) is very 
small compared to the $|a\rangle$ peaks (orbital sate \#1), clearly 
these two rates must be very different.  As $B$ is swept through 
the avoided crossing region, the admixture of the two basis states 
within the eigenstates changes, with $|\gamma(B)|$ evolving 
gradually from 1 to 0, and $|\eta(B)|$ going from 0 to 1.  This 
means that there should be a gradual exchange of tunneling weight 
between $|lower \rangle$ and $|upper \rangle$, with 
$\Gamma$$_{L,lower}$ evolving from $\Gamma$$_{L,a\uparrow}$ 
to $\Gamma$$_{L,b\downarrow}$, and $\Gamma$$_{L,upper}$ doing 
the reverse.  Qualitatively, this crossover behavior is apparent in 
the currents in Fig.~2(b).

	In order to attempt a more quantitative treatment of the 
measured currents, it is necessary to analyze the relationship 
between the bare tunneling rates $\Gamma$ discussed above, and 
the value of the current that results from sequential tunneling 
across the two tunnel barriers.  This requires a full consideration of 
all the processes that can occur during current flow.  When the 
applied bias is larger than the level spacing, non-equilibrium 
electron distributions are produced within the nanoparticle during 
tunneling, and these can open new channels for electrons to flow 
\cite{Agam1,noneq}.  The idea is shown in Fig.~4.  In Fig.~4(a) 
we show the simple process of an electron tunneling from the left 
electrode to an empty level on the particle, when the voltage across 
the device is sufficient to supply the threshold tunneling energy.  
Due to electrostatic interaction with this additional electron upon 
its arrival, the lower-energy electronic states already filled within 
the particle can be shifted up in energy to the positions drawn.  If 
the applied voltage needed to initiate tunneling is larger than the 
level separation between
\linebreak
\begin{figure}
\vspace{-1.1cm}
\begin{center}
\leavevmode
\epsfxsize=2.1in
\epsfbox{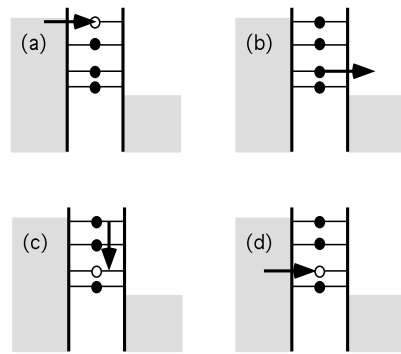}
\end{center}
\caption{
(a)-(d) Some of the allowed transitions contributing to the 
magnitude of the current flowing at a tunneling threshold.}
\end{figure}
\noindent
states, one possibility for the next step in 
the tunneling process might be as shown in Fig.~4(b), where an 
electron tunnels out of one of these lower-energy states to the right 
electrode, leaving an electron-hole excitation on the particle.  After 
this, the excited electron might relax (Fig.~4(c)) before the next 
electron tunnels onto the particle.  Alternatively an electron may 
tunnel from the left electrode into the hole (Fig.~4(d)).  All of these 
processes, and all other energetically-allowed transitions, will 
contribute to the value of the current that results when the voltage 
is turned above the threshold required to initiate tunneling in 
Fig.~4(a).  The lower-energy, initially-filled states can contribute 
to the current even though they are hidden in the sense that they do 
not produce tunneling thresholds of their own.  In general, to 
calculate the total current, one must solve a full master equation 
which takes into account the rates of all the allowed transitions, 
including the hidden levels.  The parameters entering the 
calculation are the number of hidden levels, the tunneling rates 
between each level and the left and right leads, and the relaxation 
rates for the different allowed excitations within the particle.

	For samples of the sort investigated in Figs.~1-3, which do 
not have a gate electrode that can be used to adjust the number of 
hidden states, the measured data are not sufficient to fully 
determine all of the parameters required to describe the currents 
quantitatively.  However, progress can be made with some 
simplifying assumptions.  One particular difficulty is that we do 
not know how the relaxation rate for excitations within the particle 
({\em e.g.}, Fig.~4(c)) compares to the tunneling rates in this 
sample.  The predicted order of magnitude for the relaxation rate 
due to electron-phonon scattering is $10^8 s^{-1 }$ \cite{Agam1}, 
less than the tunneling rates we will determine, but only by about a 
factor of 10.  Allowing for some uncertainty in the theory, we will 
consider both of the two simple limits -- that the relaxation rate is 
either much slower or much faster than the tunneling rate.  The 
following discussion should not be considered a quantitative 
determination of tunneling parameters, but it will serve to illustrate 
the way in which the measured changes in current increments as a 
function of $B$ can be linked to the bare tunneling rates.  We also 
invoke three other simplifying assumptions: (1) that the number of 
hidden states stays the same over the range of voltage and 
magnetic field analyzed in Figs.~1(b) and 2, (2) that the ratios 
$\Gamma_{L,i} / \Gamma_{R,i}$, for the tunneling rates from 
quantum level $i$ to the left and right electrodes, are all the same, 
and (3) that for all {\em hidden} states the tunneling rates to the 
left electrode are identical $\equiv \Gamma_{L,h}$.  We define 
$x\!=\!\Gamma_L / \Gamma_R$.  These assumptions reduce the 
free parameters in the problem to a tractable number, but we make 
no rigorous claims as to plausibility.  

	We first consider the slow-relaxation limit, in which we can 
ignore all processes of the sort pictured in Fig.~4(c).  In this limit, 
the second assumption listed above leads to a great simplification, 
because the probability for occupation of any quantum level 
accessible by tunneling will be the same.  If the voltage bias is 
such that $N$ hidden levels and $M$ originally-empty levels 
participate in tunneling, the total current that results at $T\!=\!0$ is, 
by solution of an elementary set of rate equations, for positive 
$(+)$ and negative $(-)$ bias:
\begin{equation}
\matrix{\displaystyle I_{M,N,+} = {{eM(N+1) (N \Gamma_{L,h} + 
\sum_{i\!=\!1}^{M} \Gamma_{L,i})} \over {(N+M) (N+1+xM)}} 
\cr
\displaystyle I_{M,N,-} = {{eM(N+1) (N \Gamma_{L,h} + 
\sum_{i\!=\!1}^{M} 
\Gamma_{L,i})} \over {x(N+M) (N+1+{{M} \over {x}})}} }
\end{equation}
Within the model, we can determine the values for the 4 free 
parameters $N$, $x$, $\Gamma$$_{L,h}$, and $\Gamma$$_{L,1}$ 
(the tunneling rate into the lowest-energy initially-empty orbital 
state at low field) from the four low-field ($B\!=\!2$ Tesla) values 
of the total, cumulative currents flowing at the $1\uparrow$ and 
$1\downarrow$ thresholds, for positive and negative bias: 
$I_{1,N,+}= \Delta$$I_{1\downarrow,+} = 9.6\times 
10^{-11}$~A, $I_{2,N,+}= \Delta$$I_{1\uparrow,+} + 
\Delta$$I_{1\downarrow,+} = 1.41\times10^{-10}$~A, $I_{1,N,-}
= \Delta$$I_{1\downarrow,-} = 6.6\times10^{-11}$~A, and 
$I_{2,N,-}= \Delta$$I_{1\uparrow,-} + \Delta$$I_{1\downarrow,-} 
= 1.19\times10^{-10}$~A.  Because of time reversal symmetry 
we can assume that $\Gamma$$_{L,1\uparrow} = 
\Gamma$$_{L,1\downarrow}$ for small $B$.  The results for the 
four free parameters are $N = 2.4$, $x = 2.0$, $\Gamma$$_{L,h} 
=9.4 \timesÊ10^8Ês^{-1}$, and $\Gamma$$_{L,1} = 1.01 \times 
10^9Ês^{-1}$.  The fact that $N$ is not an integer may reflect the 
weaknesses of the assumption that the ratio $x = 
\Gamma$$_L/\Gamma$$_R$ is the same for all the quantum levels 
and/or the assumption of slow relaxation.  Employing these values 
and the measured (positive-bias) current increments for the 
avoided-crossing states shown in Fig.~2(b), we can then invert 
Eq.~(11) (for $M=$2 and 3) to estimate the bare tunneling rates 
$\Gamma$$_{L,lower}$ and $\Gamma$$_{L,upper}$ over the whole 
range of $B$ from 1 to 7~Tesla, with the results shown in 
Fig.~3(b).  

	In the same way we can consider the fast-relaxation limit, 
in which the electrons in the nanoparticle relax to their lowest 
energy state between all tunneling events.  The solutions to the rate 
equations are
\begin{equation}
\matrix{\displaystyle I_{M,N,+} = {{e (\Gamma_{L,1}+N 
\Gamma_{L,h})(\sum_{i\!=\!1}^{M} \Gamma_{L,i})} \over 
{\Gamma_{L,1}+N \Gamma_{L,h}+x(\sum_{i\!=\!1}^{M} 
\Gamma_{L,i})}} \cr
\displaystyle I_{M,N,-} = {{e (\Gamma_{L,1}+N 
\Gamma_{L,h})(\sum_{i\!=\!1}^{M} \Gamma_{L,i})} \over 
{x\lbrack \Gamma_{L,1}+N \Gamma_{L,h}+(\sum_{i\!=\!1}^{M} 
\Gamma_{L,i})/x}}. }
\end{equation}
In this case there are just three parameters, $\Gamma_{L,1}$, $x$, 
and $N \Gamma$$_{L,h}$, which can be determined from the 
$B\!=\!2$~Tesla values of $I_{1,N,+}= 
\Delta$$I_{1\downarrow,+}$, $I_{2,N,+}= 
\Delta$$I_{1\uparrow,+} + \Delta$$I_{1\downarrow,+}$, and 
$I_{1,N,-}= \Delta$$I_{1\downarrow,-}$ as $\Gamma_{L,1} =9.3 
\times 10^8 s^{-1}$, $x = 1.96$, and $N \Gamma$$_{L,h} = 2.4 
\times 10^9 s^{-1}$.  These parameters, together with Eq.~(12) 
predict a value of $1.18 \times 10^{-10}$~A for $I_{2,N,-}= 
\Delta$$I_{1\uparrow,-} + \Delta$$I_{1\downarrow,-}$, in good 
agreement with the measured value, $1.19 \times 10^{-10}~A$.  We 
can then invert Eq.~(12) using the measured current increments of 
the avoided-crossing states in Fig.~2(b) to estimate the tunneling 
rates $\Gamma_{L,lower}$ and $\Gamma_{L,upper}$ in the fast 
relaxation limit (Fig.~3(c)).  The differences between Figs.~3(b) 
and 3(c) reflect to some extent the degree of uncertainty with 
which we can estimate these bare tunneling rates.

	We see from both Figs.~3(b) and (c) that the crossover 
observed in the magnitude of the current increments for the two 
avoided crossing states can be related to a crossover in the bare 
tunneling rates, of the type predicted by the spin-orbit scattering 
Hamiltonian.  The lines in Fig.~3 (b,c) display fits to the S-O 
model result, Eq.~(9), with only 2 adjustable parameters, 
$\Gamma$$_{L,a\uparrow}$, and $\Gamma$$_{L,b\downarrow}$, 
which simply set the $B\!=\!0$ values of the tunneling rates.  For 
the slow-relaxation limit the $B$-dependence of the tunneling rates 
is very well described by the S-O formalism.  In particular, as 
predicted by the model, the tunneling rates cross close to the same 
magnetic field value, 5.4~Tesla, where the avoided-crossing levels 
have their closest approach.  Also, the tunneling rates well beyond 
the crossover regime are approximately equal to the $B\!=\!0$ 
tunneling rates.  Neither result holds for the current increments 
themselves (Fig.~2(b)), due to the effect of the hidden levels.  The 
agreement between the S-O theory and the tunneling parameters 
estimated in the fast-relaxation limit is not quite as close as for the 
slow-relaxation limit.  This is consistent with the estimates in 
ref.~\cite{Agam1} that the energy relaxation rate is $10^8 s^{-1}$, 
an order of magnitude less than the tunneling rates we determine.

\vskip 24pt
\centerline{III.  Effects of Spin-Orbit Interactions on}
\centerline{Superconducting Nanoparticles}
\vskip 12pt
\nobreak

	We next consider different samples, a larger Al particle 
containing 4\% Au impurities (Fig.~5(a)), compared to a pure Al 
particle of similar size showing no indications of S-O scattering 
(Fig.~5(b)) \cite{samedata}.  Because of their larger size, the mean 
level spacings in both samples are smaller than in the device of 
Figs.~1-3, but nevertheless a large energy difference is visible 
between the first and second
peaks in both spectra.  This is 
characteristic of odd-to-even tunneling in a superconducting 
particle.  The energy
\linebreak
\begin{figure}
\vspace{-1.1cm}
\begin{center}
\leavevmode
\epsfxsize=2.8in
\epsfbox{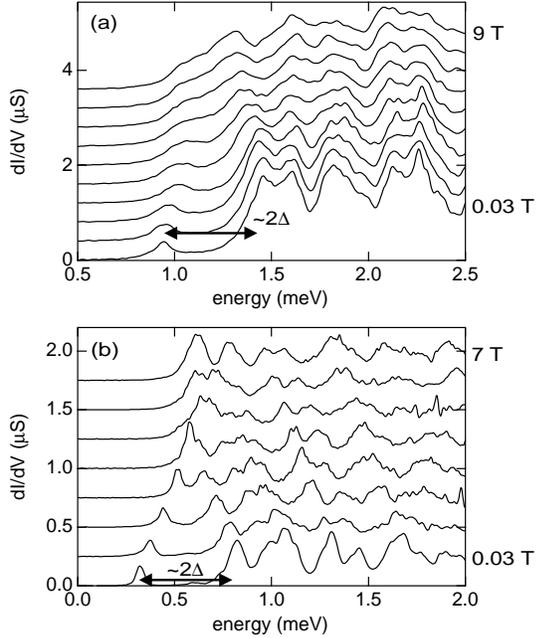}
\end{center}
\caption{
(a)  Tunneling spectrum for an Al particle containing 4\% 
Au impurities, for a sequence of magnetic fields from 0.03 
to 9~Tesla in 1~Tesla increments, $T\!=\!15$ mK.  The 
particle exhibits a superconducting gap for odd-to-even 
electron tunneling.  (b) Comparison data for a pure Al 
particle, from ref.~[5].  The curves are artificially offset for 
visibility.}
\end{figure}
\noindent
gap reflects the large difference in energy 
(approximately twice the superconducting gap $\Delta$) required 
for the tunneling of an electron to reach the ground state of an 
even-electron superconductor in which all electrons are paired, 
versus the first excited state containing two unpaired quasiparticles 
\cite{cbt}.  The two samples display qualitatively different 
behavior in a number of respects, however.  One difference is that 
the resonance peaks in the Au-doped sample are somewhat 
broader.  We believe that this is {\em not} related to the impurities, 
but is instead an effect of non-equilibrium distributions of electrons 
on the island \cite{Agam1,Agam2}, excited by a source-drain 
voltage 3 times as larger than what is needed to overcome the 
Coulomb blockade and initiate tunneling in Fig.~5(a).  Similar 
broadening could also be observed for the particle in Fig.~5(b) (see 
ref.~\cite{rbt2}, Fig.~4), when a gate voltage was used to shift the 
tunneling spectrum to comparable values of the source-drain 
voltage.  The device in Fig.~5(a) had no gate.  

	We will focus instead on the differences in the magnetic 
field dependence of the data in Fig.~5(a) and (b).  For the pure Al 
superconducting nanoparticle, the primary effect of a magnetic 
field is to produce linear shifts corresponding to a 
Zeeman-spin-splitting 
with $g_{\rm{eff}} = 2 \pm 0.05$ \cite{wrongg}.  The gap in 
the spectrum decreases linearly due to the difference in Zeeman 
energies for the ground and first-excited states of the 
superconductor, until it goes to zero at about 4~Tesla.  Models of 
superconductivity in small particles \cite{Braun1,Braun2} relate 
this crossing with the superconducting critical field, because the 
extent of electron pairing correlations drops abruptly at this point, 
although fluctuation-induced effects of attractive electron 
interactions may persist \cite{Matveev}.  In contrast, all the 
resonance energies for the sample containing gold impurities are 
significantly less sensitive to an applied magnetic field.  Instead of 
Zeeman splitting with $g_{\rm{eff}}\!=\!2$, the ground and 
first-excited-state transitions move at low $B$ with slopes 
$g_{\rm{eff},1}/2+ g_{orb,1} = 0.41 \pm 0.03$ and $-g_{\rm{eff},2}/2+ 
g_{orb,2} = -0.27 \pm 0.03$, suggesting values for $g_{\rm{eff}}$ in 
the range 0.5-0.8.  Even at 9~Tesla, the gap between these states 
has not gone to zero, indicating a much larger critical field for 
superconductivity in this sample than for pure Al.  Similar 
increases in critical fields due to the reduction in the effective 
$g$-factor caused by S-O scattering are familiar for thin films in 
parallel magnetic fields, and in other contexts where 
superconductivity is limited by spin-induced pair-breaking 
\cite{Clogston}.  At fields above 6~Tesla, the slope of the energy 
vs.~$B$ curve of the ground-state transition in the Au-doped 
sample changes sign (with the energy decreasing with increasing 
$B$ at high fields), suggesting an avoided crossing with the 
higher-lying levels.  The minimum gap between the ground-state 
and first excited-state peaks corresponds to a S-O matrix element 
of magnitude approximately 130 $\mu$$eV$ \cite{est2}.  

	The presence of S-O scattering must necessarily change the 
nature of the superconducting transition in a magnetic field.  As we 
noted above for pure Al particles, the extent of superconducting 
pairing correlations is predicted to drop abruptly at the magnetic 
field for which the energy of the first state that moves to lower 
energy with increasing $B$ (meaning that it is a spin-1 tunneling 
state) crosses below the energy of the upward-trending (spin-0) 
ground state, so that it becomes energetically favorable to break a 
Cooper pair.  In contrast, in the particles with significant S-O 
scattering, the existence of avoided crossings means that energy 
levels corresponding to different spin states do not cross.  
Therefore the disruption of pairing correlations must occur 
gradually, as the spin content of the particle's ground state changes 
continuously in the avoided crossing region.

	Notably, the magnitude of the superconducting gap at 
$B\!=\!0$ is not significantly affected by the presence of S-O 
scattering.  Setting the difference between the ground and 
first-excited state energies equal to $2\Delta$, 
we have $\Delta \approx$ 0.25 
meV for the pure Al particle of Fig.~5(b), similar to previous 
values \cite{prevval}, and $\Delta \approx$ 0.26 meV for the 
particle with Au impurities.  This similarity is as expected, since 
S-O scattering does not break time-reversal symmetry and therefore 
does not interfere with superconducting pairing.
\vfill\eject

\vskip 24pt
\centerline{IV.  Conclusions}
\vskip 12pt
\nobreak
	We have examined a number of effects associated with the 
presence of S-O scattering in metal nanoparticles.  The sensitivity 
of the ``electron-in-a-box'' energy levels to an applied magnetic 
field is weakened, so that they exhibit effective $g$-values less 
than 2.  When predominantly spin-up and spin-down levels 
approach each other as a function of magnetic field, they may 
undergo avoided crossings due to S-O-induced coupling between 
the spin-up and spin-down states.  In the avoided crossing region, 
the magnitude of the current transported at the resonance 
thresholds can change, due to the changing admixture of spin-up 
and spin-down basis states that comprise the energy eigenstates.  
The presence of Au impurities does not greatly modify the size of 
the superconducting gap in Al particles large enough to exhibit 
superconductivity.  However, the critical magnetic field for the 
destruction of superconductivity is increased.  We also argue that 
with the presence of S-O scattering, the superconducting pairing 
parameter should vary continuously at large fields, because S-O 
scattering eliminates the simple level crossings which cause the 
extent of pairing correlations to drop abruptly in pure Al samples.  
As a final remark, we note that all of the results that we describe 
can be adequately explained by treating the S-O interaction 
perturbatively, and by ignoring the effect of the magnetic field on 
electron orbits.  For samples with stronger S-O interactions or with 
larger sizes such that effects of an applied field on the orbital states 
are significant, a more sophisticated treatment would be necessary 
\cite{Glazman}.

	We thank Leonid Glazman, Konstantin Matveev, and Jan 
von Delft for valuable discussions.  The work at Cornell was 
supported by the MRSEC program of the NSF (DMR-0632275), 
the Sloan Foundation, and the Packard Foundation, and was 
performed in part at the Cornell Nanofabrication Facility, funded 
in part by the NSF (ECS-9319005), Cornell University, and 
industrial affiliates.  The work at Harvard was supported by NSF 
Grant No. DMR-92-07956, ONR Grant No. N00014-96-1-0108, 
JSEP Grant No. N00014-89-J-1023, and ONR AASERT Grant No. 
N00014-94-1-0808.

\end{multicols}

\begin{thebibliography}{99}
\bibitem{estimate} The spin-orbit scattering rate of interest for 
samples with a continuum density of states, $1/\tau_{SO}$ may be 
estimated [10] in terms of a mean-squared 
spin-orbit matrix element and the mean level spacing, $\delta$
\begin{equation}
{{1} \over {\tau_{SO}}} \sim {{2\pi} \over {\hbar \delta}} 
\overline{|\langle H_{SO} \rangle |^2}.
\end{equation}
For the data discussed in Figs.~1-3, the order of magnitude for this 
quantity is $10^{10}~s^{-1}$.
\bibitem{Zerob}  We do not display $B\!=\!0$ data, because in 
this case the aluminum electrodes are superconducting, and this 
shifts the resonance peaks.
\bibitem{Upbound}  An experimental upper bound on the shift in 
the average energies of Zeeman-split peaks in this sample would 
correspond to an effective orbital $g$-factor less than 0.1.  We are 
not able to comment on the functional form of any dependence of 
the orbital energy on magnetic field.
\bibitem{noneq}We will ignore shifting of the single-particle 
tunneling energies which may be produced by non-equilibrium 
electron occupations, producing clusters of tunneling resonances 
[6,7], because these features are not resolved in this device.
\bibitem{samedata} The data in Fig.~5(b) is from the same sample 
as the data in Figs.~ 2-4 of ref.~[5].
\bibitem{wrongg} Claims of smaller $g$-factors made in ref.~[4] 
are wrong, the result of confusing different orbital states as 
Zeeman-split spin states.  This was made clear in ref.~[5], where 
we observed that upward-trending Zeeman states can have 
significantly smaller amplitude than downward-trending states, 
making them difficult to observe.
\bibitem{est2} The estimate for this matrix element would be 
decreased if the lower energy state is influenced by coupling to 
several higher-energy states in the avoided crossing region.
\bibitem{prevval} Values quoted previously in [4,5] for $\Delta$ 
in pure Al particles, 0.29-0.38 meV, were determined in a slightly 
different way -- by taking the difference between the low-$B$ 
threshold for tunneling and the average threshold at high field 
where superconductivity is suppressed.  Estimating $2\Delta$ from 
the difference in energy between the 2-quasiparticle state and the 
fully-Cooper-paired state in an even-electron particle is expected to 
give a slightly smaller answer for $\Delta$, because of gap 
suppression in the 2-quasiparticle state.
\end{thebibliography}
\end{document}